\documentstyle[aps,prl,multicol,epsfig]{revtex}

\begin{document}

\title{Anomalous resonance phenomena of solitary waves with internal modes}

\author{Niurka R.\ Quintero$^{*}$ and Angel S\'anchez$^{\dag}$}

\address{Grupo Interdisciplinar de Sistemas
Complicados (GISC), Departamento de Matem\'aticas,
Universidad Carlos III de Madrid,\\
Avenida de la Universidad 30, E-28911 Legan\'{e}s, Madrid, Spain} 

\author{Franz G.\ Mertens$^{\ddag}$}

\address{Physikalisches Institut, Universit\"at Bayreuth,
D-95440 Bayreuth, Germany}
              
\date{\today}

\draft

\maketitle

\begin{abstract}    
We investigate the non-parametric, pure ac driven dynamics of 
nonlinear Klein-Gordon solitary waves 
having an internal mode of frequency $\Omega_i$. 
We show that the strongest resonance 
arises when the driving frequency $\delta=\Omega_i/2$, whereas when 
$\delta=\Omega_i$ the resonance is weaker, disappearing  
for nonzero damping.
At resonance, the dynamics of the kink center of mass becomes chaotic. 
As we identify the resonance mechanism
as an {\em indirect} coupling to the internal mode due to its symmetry, 
we expect similar results for other systems.
\end{abstract}    

\begin{multicols}{2}
\narrowtext

An important paradigm established over the last two decades is
that {\em solitary waves} 
or {\em solitons} 
behave very much like point particles when subjected to (a large class 
of) external forces and perturbations \cite{Scottbook,Review,Yura3}. 
However, many solitary waves possess
one (sometimes, more than one) {\em internal} or {\em shape mode} 
\cite{Michel,David}, and in 
that case the particle picture of their dynamics may be
oversimplified:
Indeed, the internal mode can temporarily store energy and release it at a 
later stage, giving rise to resonance phenomena in solitary wave 
collisions \cite{David} or in solitary wave interactions with 
inhomogeneities \cite{Yura}. As 
internal modes are quite common in nonlinear systems,
either intrinsically  or a result of
small perturbations \cite{Yura2}, the question of their influence on the
dynamics of solitary waves is a very general and relevant one.

One aspect of solitary wave dynamics that 
has proven itself difficult to understand 
is that of {\em topological solitary waves} or 
{\em kinks} subjected to pure, i.e., {\em non-parametric ac driving}. 
Only recently \cite{Niurka} 
the ac driven dynamics of sine-Gordon kinks has been definitely clarified, 
settling a question that had received different and contradictory answers.
However, since the sine-Gordon kink does not have an internal 
mode, it does not provide information about the interplay 
of those modes with external ac drivings, which, to our knowledge, 
has only been analyzed for 
{\em breathers}, 
non-topological excitations with an internal mode 
(see \cite{Niurka} and references therein). Naively, 
the only new phenomenon one expects from a non-parametric external
driving is a resonance when its frequency, 
$\delta$, matches that of the internal mode, $\Omega_i$. 
The aim of this letter is to show that, in
fact, the actual scenario is most unexpected and highly non-trivial. 
As we will see below, {\em a strong, anomalous resonance arises when 
$\delta=\Omega_i/2$, whereas the normal resonance at $\delta=\Omega_i$ 
is definitely weaker, only possible at exactly zero damping, and even
then it can be suppressed} by appropriate choices of other parameters. 
We expect this result to be generic, because our analytical
approach allows us to identify the mechanism for such a 
peculiar phenomenon: The ac force does not act
{\em directly} on the internal mode (because of {\em symmetry} reasons), 
but rather, they interact {\em indirectly} via the
translational motion which couples to the internal mode. Our predictions
are fully confirmed by numerical simulations, which in addition
show 
the implications of these resonances for the kink dynamics. 

As a specific example of a kink with internal mode, we take the 
well known \cite{Scottbook} $\phi^4$ equation, which, when driven with 
an ac force  $f(t)=\epsilon \sin(\delta t + \delta_{0})$, reads
\begin{equation}
\phi_{tt} - \phi_{xx} +U'(\phi) = -\beta \phi_{t} + f(t),
\label{ecua1}
\end{equation}
where $U(\phi)=(\phi^2-1)^2/4$ and $\beta$ is a damping coefficient. 
Previous related works on this system are \cite{Gonzalez}, where
resonances in the 
presence of an external (time independent) potential have been considered,
and \cite{Yura4}, that dealt with non-resonant, high frequency
parametric ac drivings. For our problem, 
our analytical approach will be the well known 
{\em collective coordinate} (CC) method \cite{Review,Yura3}. 
A first order of approximation is given by the 
McLaughlin-Scott method \cite{McL}:  We assume that
the solution of (\ref{ecua1}) is of the form 
\begin{equation}
{\displaystyle{\phi(x,t)=\tanh 
\Big [\frac{x - X(t)}{l_0 \sqrt{1-V(t)^2}} \Big ]}},
\label{ecua2}
\end{equation}
where $l_0=\sqrt{2}$. 
The center of the kink $X(t)$ and its velocity $V(t)$ are related
by ${X(t)=\int_{0}^{t} dt' V(t')+X(0)}$, and both
are unknown functions describing the motion of the kink as a coherent
entity. By means of a standard procedure \cite{Review,Yura3,Niurka,McL} 
involving conservation laws, 
an ordinary differential equation of motion 
for $V(t)$
can be obtained, 
linearized \cite{Niurka} and 
solved, yielding 
\begin{eqnarray}
\label{ecua3}
V(t) & = & \frac{r(t)}{\sqrt{1+r(t)^2}},\\
r(t) & \equiv & \bar{c} \, e^{-\beta t} -
\frac{3 \sqrt{2} \epsilon}{2 (\beta^{2} + \delta^{2
})} \times \nonumber \\
\nonumber
& & \times [\beta \sin(\delta t + \delta_{0}) - \delta \cos(\delta t +
\delta_{0})],\\
\bar{c}&=&\gamma_0{V(0)} + \frac{3 \sqrt{2} \epsilon}
{2 (\beta^{2} + \delta^{2})}
[\beta \sin(\delta_{0})- \delta \cos(\delta_{0})],
\label{ecua5}
\end{eqnarray}
where $\gamma_0\equiv 1/{\sqrt{1-V(0)^{2}}}$.
For the undamped case 
($\beta = 0$),
we see that the kink oscillates if
$\gamma_0{V(0)} = {3 \sqrt{2} \epsilon}\cos(\delta_0)/({2 \delta})$;
otherwise, the kink will move either
to the right or to the left, depending on the relation between the parameters
of the ac force and the initial velocity. Note also that such a dc
motion is absent when $\beta \ne 0$. These results were numerically
confirmed for sine-Gordon kinks (which do not have internal modes) in 
\cite{Niurka}.     

In order to include
internal mode effects, we proceed as follows: We rewrite Eq.\ (\ref{ecua1})
as 
\begin{equation}
\label{ecua6}
\dot{\psi} = -\frac{\delta {\it {H}}} {\delta \phi} -
\beta \dot{\phi} + f(t),\quad\quad\quad 
\dot{\phi}  =  \frac{\delta {\it {H}}} {\delta \psi},
\end{equation}
where $\psi = \dot{\phi}$, the dot meaning derivative with respect to time, 
 and
\begin{eqnarray}
{\it{H}} = \int_{-\infty}^{+\infty} dx \Big \{
\frac{1}{2} \psi^2 + \frac{1}{2} \phi_{x}^{2} + U(\phi) \Big\},
\label{ecua8}
\end{eqnarray}
is the Hamiltonian of the system when
$\epsilon=\beta=0$.
We now make the {\em Ansatz}\/ 
$\phi(x,t) = \phi[x - X(t), l(t)],$
whereas from the definition of $\psi$ we have that
$\psi(x,t) = \psi[x - X(t), l(t), \dot{X}, \dot{l}]$. 
As in the McLaughlin-Scott method,
$X(t)$ represents the kink center position, but now
we introduce a second collective variable $l(t)$, that
will stand for the kink width, i.e., the internal 
mode excitation, below.

The procedure to obtain the CC equations corresponding
to this generalized travelling wave {\em Ansatz}\/ has been put forward in
\cite{franza,franz}. Basically, it consists of inserting our {\em Ansatz}\/
into (\ref{ecua6}), multiplying the first equation 
by ${\partial \phi}/{\partial X}$ 
and the second one by 
${\partial \psi}/{\partial X}$ (${\partial
\phi}/{\partial l}$ and ${\partial \psi}/{\partial l}$), 
taking their difference
and integrating over $x$. This yields
a rather cumbersome, ordinary differential equation for $X(t)$ [$l(t)$],
which we omit here for brevity.
The next step is to choose a specific functional form for $\phi$, which we
do following the work of Rice \cite{Rice}, and let 
\begin{eqnarray}
\phi[x-X(t), l(t)] =
\phi_0 \Big [\frac{x - X(t)}{ l(t)} \Big],
\label{ecua17}
\end{eqnarray}   
where $\phi_0[(x-X)/l_0]$  is the static kink solution, which is 
and odd function
(with respect to its center) for the $\phi^4$ and for any other 
even potential.
Upon particularization of the CC equations
for this form for $\phi$ we finally obtain
\begin{eqnarray}
M_{0} l_{0} \frac{\ddot X}{l} - M_{0} l_{0} \frac{\dot X \dot l}{l^{2}}
 &=&  -  \beta M_{0} l_{0} \frac{\dot X }{l} - 2 f(t),
\label{ecua18} \\
\alpha M_{0} l_{0} \frac{\ddot l}{l} + M_{0} l_{0} \frac{\dot X^{2}}{l^{2}}
 &=  &K^{int}(l,\dot{l},\dot{X}) -  \beta \alpha M_{0} l_{0} \frac{\dot l}{l};  \label{ecua19} 
\end{eqnarray}
where
$K^{int}  =  -{\partial E}/{\partial l}$ and 
\begin{equation}
E =\frac{1}{2} \frac{l_{0}}{l} M_{0}  \dot{X}^{2}+ 
\frac{l_{0}}{2l} \alpha M_{0}  \dot{l}^{2} +
\frac{1}{2} M_{0} \Big(\frac{l_{0}}{l} + \frac{l}{l_{0}}\Big)
\label{ecua21}
\end{equation}
is the kink energy. Importantly, in obtaining 
Eq.\ (\ref{ecua19}) a term of the 
form $f(t)\int_{-\infty}^{\infty}dx\,\partial\phi_0/\partial l$, coming
{}from the coupling of the ac driving to the internal mode, has vanished 
because of symmetry. For the $\phi^4$ equation, $\phi_0=\tanh x$ in 
Eq.\ (\ref{ecua17}), 
which yields 
$\alpha=(\pi^2-6)/12$ and $M_{0}=2 \sqrt{2}/3$. 

Let us now simplify these expressions in order to make its physical 
significance more transparent. To begin with, Eq.\ (\ref{ecua18}) 
can be solved for $\dot{X}/l$, yielding: 
\begin{eqnarray}
P&\equiv&\frac{M_{0} l_{0}\dot{X}}{l}  = 
\nonumber
- {2\epsilon}
\frac{\beta \sin(\delta t + \delta_{0})- 
\delta \cos(\delta t + \delta_{0})}{(\beta^{2} +\delta^{2})}\\ 
&+ &e^{-\beta t} \Big[{2\epsilon}
\frac{\beta \sin(\delta_{0}) -
\delta \cos(\delta_{0})}{(\beta^{2} +\delta^{2})}+ \frac{M_{0} l_{0}V(0)}{l_{s}} \Big],
\label{ecua23}
\end{eqnarray}
where $l_{s}=l_{0}/\gamma_0$. 
Inserting Eq.\ (\ref{ecua23}) into
Eq.\ (\ref{ecua19}), we find
\begin{eqnarray}
\alpha [\dot{l}^{2} - 2 l \ddot{l} - 2 \beta l \dot{l}] & = &
\frac{l^{2}}{l_{0}^{2}} \Big [1 +  \frac{P^{2}}{M_{0}^{2}} \Big ] - 1.
\label{ecua24a}
\end{eqnarray}  
Although Eqs.\ (\ref{ecua23}) and (\ref{ecua24a}) are quite complicated,
choosing $V(0)$ and the phase $\delta_0$ so that the exponential terms in
Eq.\ (\ref{ecua23}) vanish, we 
can use a change of variables (proposed in \cite{yuri} for 
the dc driving case)
to transform the Eq.\ (\ref{ecua24a}) into
a Pinney-like equation 
(see \cite{wint} and references therein), which 
can be solved in terms of Mathieu functions \cite{Niurka2} 
if $\beta=0$ (when $\beta\neq 0$ we have not been able to solve this problem
analytically). 

In any event, we do not need the analytical expressions for the solution
of the system (\ref{ecua23}), (\ref{ecua24a}) to understand the physics 
predicted by our approach. The term $P^2$ in Eq.\ 
(\ref{ecua24a}) is an oscillatory function with frequency $2\delta$ [see
Eq.\ (\ref{ecua23})]; hence, we can immediately expect
a resonance when the external frequency $\delta$ is 
{\em half} the frequency of the internal mode, $\Omega_R=1/\sqrt{\alpha}l_0$
in the Rice approximation. For the $\phi^4$ model
$\Omega_R$ overestimates
$\Omega_i=\sqrt{3/2}$ by 1.7\% \cite{Rice}.
The analytical solution for $\beta=0$ confirms this expectation,
while numerical integration of Eq.\ (\ref{ecua24a}) proves
that the behavior of $l$ is that of a resonant, damped oscillator when 
$\beta\neq 0$. 
When $\delta=\Omega_R$, inspection of Eqs.\ (\ref{ecua23}) and 
(\ref{ecua24a}) leads to the conclusion that another resonance should
be found at $\delta=\Omega_R$ {\em only if} $\beta=0$ (otherwise it is
a transient phenomenon of lifetime $\beta^{-1}$); even then, by
choosing $V(0)$ and $\delta_0$ to cancel the non-oscillatory terms in
Eq.\ (\ref{ecua23}) the resonance is completely suppressed.
Furthermore, both analytically and numerically we have 
verified that, far away from the resonances, the 
behavior of the kink center, $X(t)$, is practically the same as the one 
predicted by the McLaughlin-Scott approach, Eq.\ (\ref{ecua3}). This means
that, within the CC framework, the behavior of ac 
driven $\phi^4$ kinks is described by the McLaughlin-Scott {\em Ansatz}, 
and only for drivings close to $\Omega_R/2$ (and $\Omega_R$ 
if $\beta=0$) 
such approach fails and resonant phenomena arise. 

This far, the CC method has provided specific predictions
about the system behavior, but in order to assess their physical reality, we
must verify that they arise in the partial differential 
equation (PDE), 
Eq.\ (\ref{ecua1}). The reason for this is twofold: First, underlying
CC methods is the assumption that no (or a negligible 
amount of) radiation is generated by the perturbation, an assumption 
whose validity can only be assessed through comparison with the 
corresponding PDE. Second, even if that is the case, the CC
equations predict an unbounded growth of $l(t)$ at resonance, 
and it is difficult to understand what does that mean in physical terms
for the kink of the full PDE, whose width is controlled by the properties
of the equation. In view of this, we  have computed the numerical solution 
of the PDE (\ref{ecua1}) by using the conservative Strauss-V\'azquez scheme
\cite{stvaz} with $\Delta x=0.1$, $\Delta t=0.01$, and a total system length 
of $L=400$. 
Our initial condition was a kink at rest and $\delta_0=\pi/2$, a pair of 
values for which we should not see a resonance at $\Omega_i$. We come back
to this below.
We monitored 
the position and the velocity of the kink center as well as 
the total energy in the system, computed from the Hamiltonian (\ref{ecua8}). 
We also tried to measure directly the kink width,
but we found that it is quite complicated to estimate it from 
the numerics, this being 
the reason why we have resorted to less direct measurements. 

\begin{figure}
\epsfig{figure=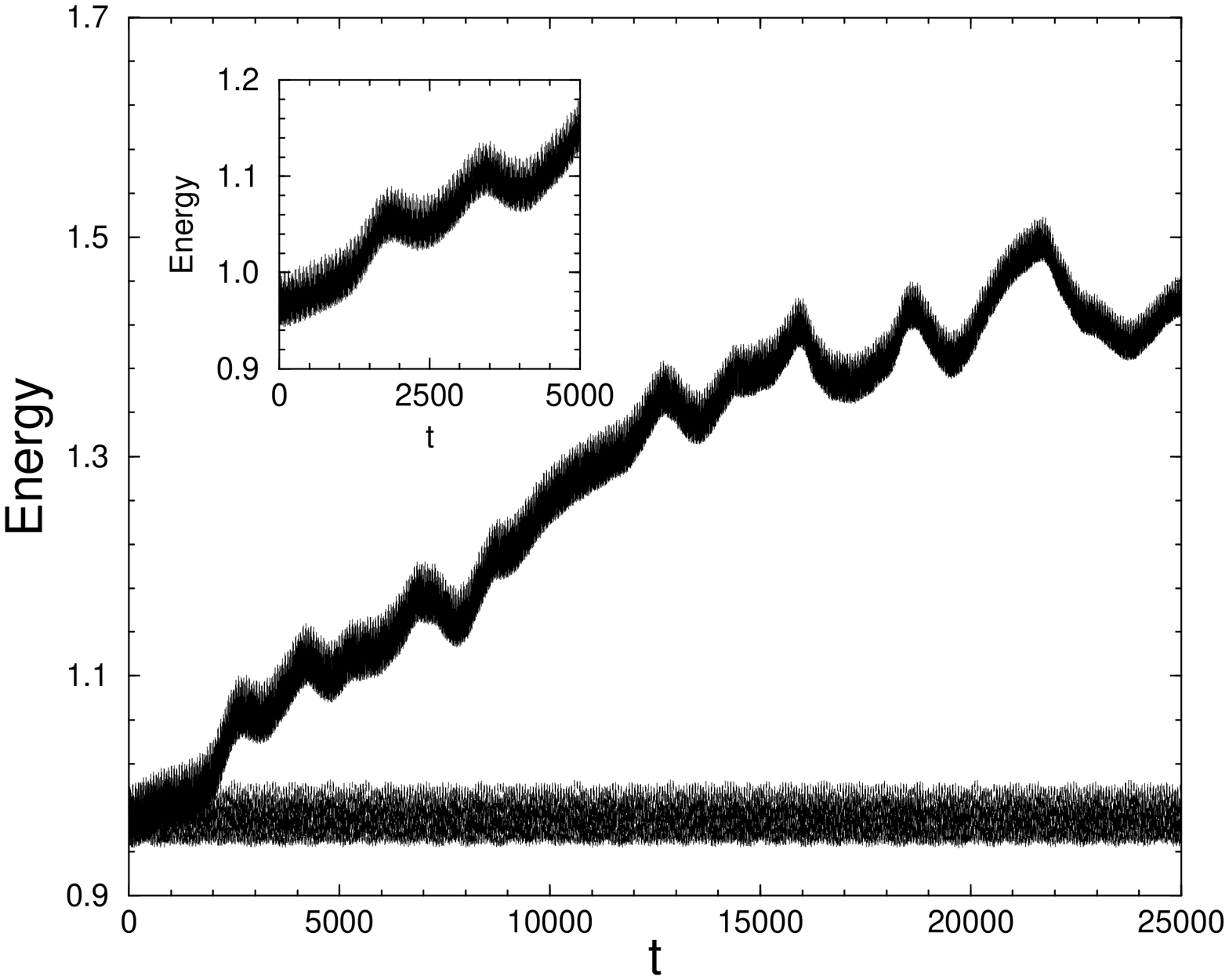,width=1.5in,height=2.8in,angle=270}

\hspace*{-.1in}\epsfig{figure=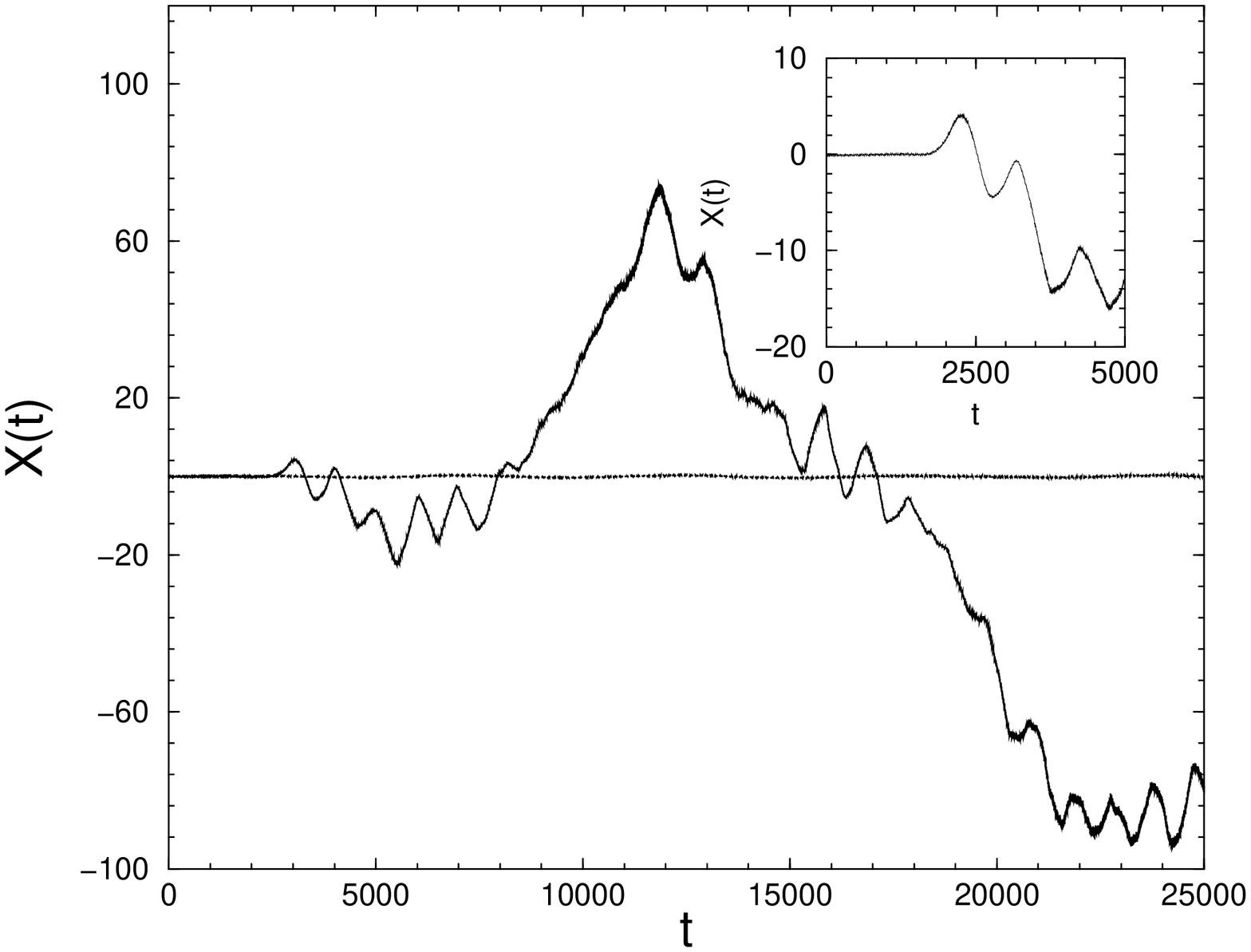,width=1.4in,height=2.9in,angle=270}
\caption[]{Results for the PDE
(\ref{ecua1}) with
$\beta=0$, $\epsilon=0.01$.
(a) Total energy
when $\delta=0.6100$ 
(upper line), $\delta=0.6080$ (lower line). (b) Same for the kink 
center, $X(t)$. Insets show the same 
for $\delta=0.6102$, a value for which the
kink reaches the boundary of the numerical system before $t=25\,000$.}
\label{fig1}
\end{figure}
Figure \ref{fig1} shows examples of the kink center dynamics and its 
energy evolution both close to and away from the predicted resonance. 
Off-resonance, the behavior of both magnitudes is 
periodic, whereas at the resonance it becomes chaotic. Specifically, the energy
increases with time: The closer to the resonance value, the faster the
increment.  The center
motion is initially periodic, until the internal mode 
amplitude has increased too much and stored too much energy, subsequently 
releasing it through its coupling with the translation mode (which we
know that exists from \cite{David}), eventually yielding the kink motion
erratic as this process is repeated once and again.
This is a clear evidence in favor of the
resonance predicted from the CC treatment. 
We have verified that at resonance the kink motion is very 
sensitive to changes in the initial conditions, hence our claim of
the appearance of chaos.

\begin{figure}
\epsfig{figure=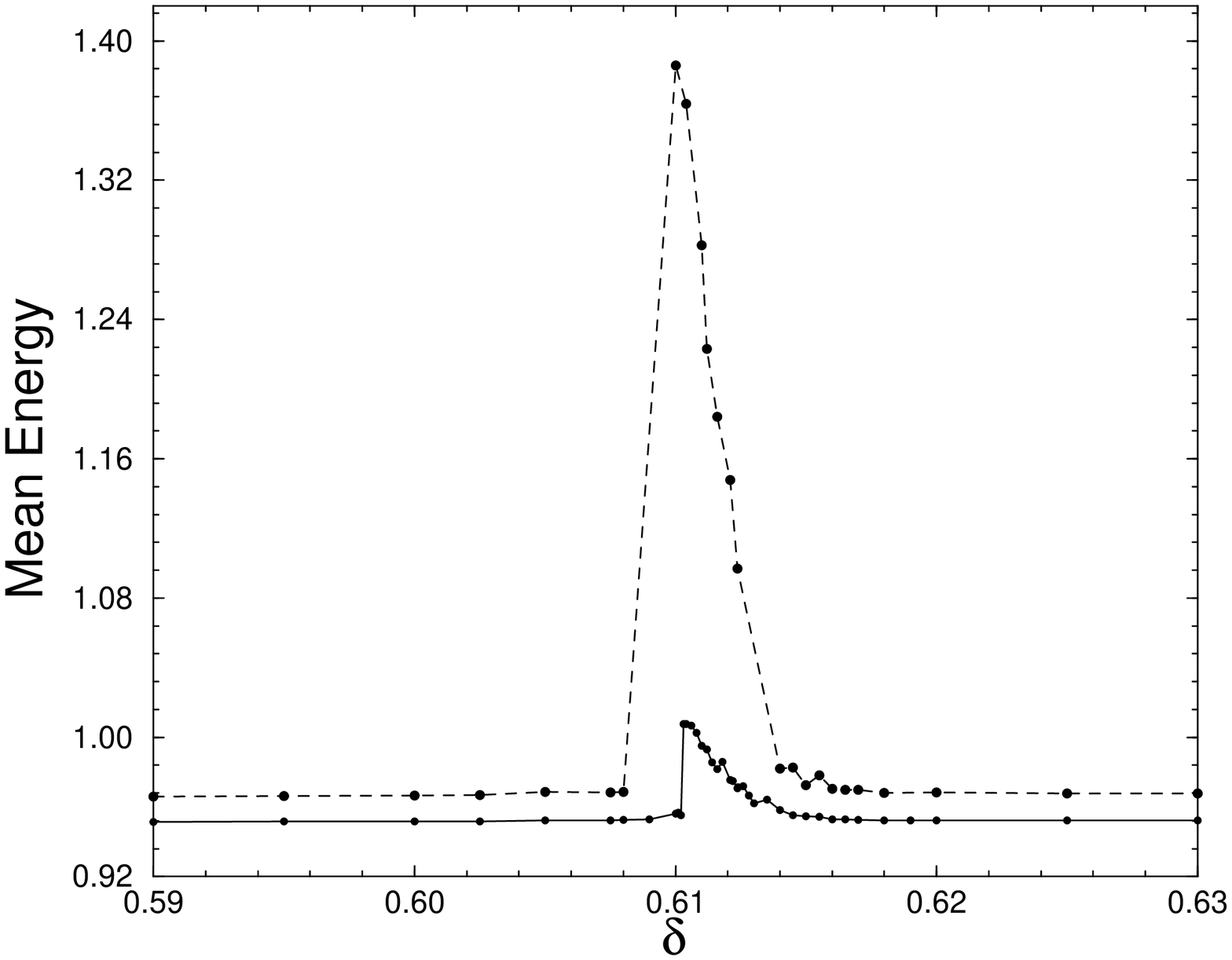,width=1.5in,height=2.8in,angle=270}

\epsfig{figure=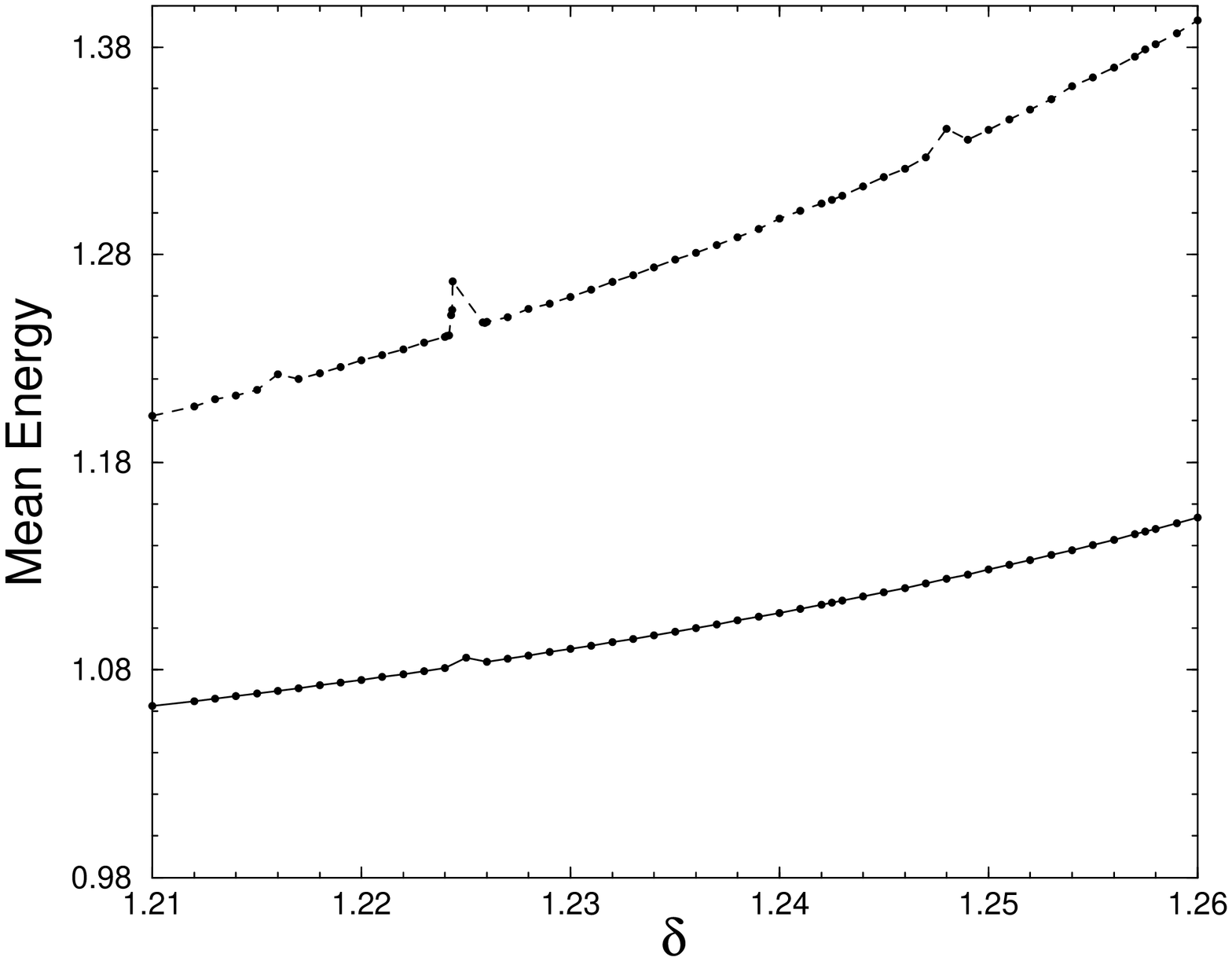,width=1.45in,height=2.8in,angle=270}
\caption[]{Results for the 
PDE (\ref{ecua1}). Mean value of the energy (see text for the way it is
computed) vs driving frequency
$\delta$, (a) close to $\Omega_i/2$, (b)
close to $\Omega_i$. In both cases, the upper 
line corresponds to $\beta=0$, and the lower line to $\beta=0.001$.}
\label{fig2}
\end{figure}
Figure 
\ref{fig2} depicts the resonance as seen through the mean 
energy, computed as a time average from $t=10\,000$ (after 
transients have died out in the damped case) to the end of the run 
at $t=25\,000$. Figure
\ref{fig2}(a) shows the behavior of this magnitude around 
$\Omega_i/2$. It is
clear from the plot that there is a strong resonance at $\delta\approx
0.6102\approx\Omega_i/2$,
both with and without dissipation. In fact, some points are missing in the 
$\beta=0$ line
because
the corresponding kinks, in their apparently random motion,
left the system before the end of the
run.
On the other hand, in spite of our choice of initial conditions,
for which no resonance is predicted at $\Omega_i$ for $\beta=0$,
a weak resonance can be seen at 
$\delta\approx 1.225\approx\Omega_i$
in Fig.\ \ref{fig2}(b). 
We note that the other two small peaks are spurious, since they appear,
disappear, or change location depending on the choice of the length 
for the numerical simulation. 
We believe that this discrepancy might come from the difficulty of
numerically tuning the condition for its suppression. In any event,
that would be a very special case, and the behavior we find for these
parameters is representative of  what occurs for
other choices of the initial velocity and the phase. Those give rise to 
a similar behavior, while, remarkably, 
the resonance at $\Omega_i$ is always weaker 
and narrower than that at $\Omega_i/2$ (cf.\ Fig.\ \ref{fig2}). 
We also see that the prediction that the
resonance at $\Omega_i$ is suppressed by the dissipation is also confirmed,
the small peak in the lower line stemming from the transients, which are 
not exactly zero for $10\,000\leq t\leq 25\,000$.
Another interesting remark is that 
Fourier analysis shows that the
monotonous increasing of the energy that appears in Fig.\ \ref{fig2}(b)
comes from the fact that, when $\delta\gtrsim 1.1$, 
the lowest phonon mode (with frequency $\omega_p=\sqrt{2}$) begins to 
be excited, 
the amplitude of its excitation monotonically increasing as $\delta\to
\sqrt{2}$. No evidence for 
lowest phonon mode excitation is seen for the resonance at $\Omega_i/2$;
therefore, this is indeed a phenomenon arising from the
coupling of the translation and the internal mode as predicted by the 
CC calculation. 

In summary, we have studied how ac forces affect solitary waves of 
kink type with an 
internal mode of frequency $\Omega_i$. 
Specifically, we have clearly shown that
the behavior of $\phi^4$ kinks under ac driving is very well described by
the two-variable CC theory we have developed here. 
The main feature of the non-parametrically, 
ac driven $\phi^4$ kink dynamics is that the 
strongest resonance occurs at $\delta=\Omega_i/2$, and not at the frequency
one would expect, $\delta=\Omega_i$.
We emphasize that this novel resonance phenomenon 
is totally unexpected from the knowledge of the internal 
mode frequency, and arises from the {\em indirect} interaction of the external
force with the internal mode via the translational motion.
Although our results have been 
obtained for a specific example, the $\phi^4$ equation (\ref{ecua1}), 
other models with even on-site potentials and internal modes,
such as the double 
sine-Gordon equation, for instance, will behave similarly
because a CC approach will lead to analogous results. 
The resonance found at the CC level manifests itself at the PDE level 
as erratic or chaotic (strongly dependent on the initial conditions)
motion of the kink as the kinetic energy of the
center of mass is stored into, and recovered from, the internal mode.

On closing, we note that the anomalous resonances described here are important 
on their own, as examples of the highly non-trivial behavior of nonlinear
systems and as hints about the mechanisms governing kink
dynamics. In addition, we think that this phenomenon should be very general,
in view of the recent finding \cite{Yura2} that perturbations of 
kink-bearing nonlinear
systems often lead to the development of an internal mode. As this occurs 
in the discrete sine-Gordon model 
\cite{Yura2}, a resonance like the one discussed
here could be relevant for the mode locking phenomena reported for that
system in \cite{Mario}. Finally, by using this discreteness induced 
internal mode, we point out that the resonance we find could be observed in 
experiments by using Josephson junction arrays as in \cite{Fernando}.

\smallskip

We thank Yuri Gaididei, Francisco Dom\'\i nguez-Adame,
 and Jos\'e Cuesta for discussions.
Work at GISC (Legan\'es) has been supported by 
DGESIC (Spain) grant PB96-0119. Travel between Bayreuth and Madrid has been
supported by ``Acciones Integradas Hispano-Alemanas'', a joint program of
DAAD (Az.\ 314-AI) and DGESIC.

\end{multicols}

\begin{thebibliography}{99}

\bibitem[*]{1} kinter@math.uc3m.es

\bibitem[\dag]{2} anxo@math.uc3m.es

\bibitem[\ddag]{3} franz.mertens@theo.phy.uni-bayreuth.de

\bibitem{Scottbook} A.\ C.\ Scott, {\em Nonlinear Science} (Oxford University,
Oxford, 1999). 

\bibitem{Review} 
A.\ S\'anchez and A.\ R.\ Bishop, SIAM Review {\bf 40}, 579 (1998),
and references therein. 

\bibitem{Yura3} Yu.\ S.\ Kivshar and B.\ A.\ Malomed, Rev.\ Mod.\ Phys.\
{\bf 61}, 763 (1989), and references therein. 

\bibitem{Michel} M.\ Peyrard and M.\ Remoissenet, Phys.\ Rev.\ B {\bf 26},
2886 (1982). 

\bibitem{David} M.\ Peyrard and D.\ K.\ Campbell, Physica D {\bf 9}, 33 (1983); 
D.\ K.\ Campbell, J.\ F.\ Schonfeld, and C.\ A.\ Wingate, {\em ibid.} {\bf 9}, 
1 (1983); D.\ K.\ Campbell, M.\ Peyrard, and P.\ Sodano, {\em ibid.} {\bf 19},
165 (1986). 

\bibitem{Yura} Yu.\ S.\ Kivshar, F.\ Zhang, and L.\ V\'azquez, Phys.\ 
Rev.\ Lett.\ {\bf 67}, 1177 (1991); Phys.\ Rev.\ A {\bf 46}, 5214 (1992).

\bibitem{Yura2} Yu.\ S.\ Kivshar, D.\ E.\ Pelinovsky, T.\ Cretegny, and 
M.\ Peyrard, Phys.\ Rev.\ Lett.\ {\bf 80}, 5032 (1998). 

\bibitem{Niurka} N.\ R.\ Quintero and A.\ S\'anchez, 
Phys.\ Lett.\ A {\bf 247}, 161 (1998);
Eur.\ Phys.\ J.\ B {\bf 6}, 133 (1998).

\bibitem{Gonzalez} J.\ A.\ Gonz\'alez, B.\ A.\ Mello, L.\ I.\ Reyes, and 
L.\ E.\ Guerrero, Phys.\ Rev.\ Lett.\ {\bf 80}, 1361 (1998). 

\bibitem{Yura4} Yu.\ S.\ Kivshar and A.\ S\'anchez, Phys.\ Rev.\ Lett.\
{\bf 77}, 582 (1996).

\bibitem{McL} D.\ W.\ McLaughlin and A.\ C.\ Scott, Phys.\ Rev.\ A {\bf 18},
1652 (1978). 

\bibitem{franza} G.\ M.\ Wysin, F.\ G.\ Mertens, A.\ R.\ V\"olkel, and A.\ R.\ Bishop, in {\em Nonlinear coherent structures in physics and biology}, 
edited by K.\ H.\ Spatschek and F.\ G.\ Mertens (NATO ASI Series B, 
vol.\ 329, Plenum, New York, 1994). 

\bibitem{franz} F.\ G.\ Mertens, H.\ J.\ Schnitzer, and A.\ R.\ Bishop,
Phys.\ Rev.\ B {\bf 56}, 2510 (1997). 

\bibitem{Rice} M.\ J.\ Rice, Phys.\ Rev.\ B {\bf 28},
3587 (1983). 

\bibitem{yuri} E.\ Majern\'\i kov\'a, Yu.\ B.\ Gaididei, and O.\ M.\ Braun, 
 Phys.\ Rev.\ E {\bf 52},
1241 (1995). 

\bibitem{wint} J.\ L.\ Reid and J.\ R.\ Ray, Z.\ Angew.\ Math.\ u.\ 
Mech.\ {\bf 64}, 365 (1984).

\bibitem{Niurka2} N.\ R.\ Quintero, A.\ S\'anchez, and F.\ G.\ Mertens, 
unpublished.

\bibitem{stvaz} W.\ A.\ Strauss, L.\ V\'azquez,
J.\ Comput.\ Phys.\ {\bf 28},
271 (1978).

\bibitem{Mario} P.\ J.\ Mart\'\i nez, F.\ Falo, J.\ J.\ Mazo, L.\ M.\ 
Flor\'\i a, and A.\ S\'anchez, Phys.\ Rev.\ B {\bf 56}, 87 (1997). 

\bibitem{Fernando} F.\ Falo, P.\ J.\ Mart\'\i nez, J.\ J.\ Mazo, and 
S.\ Cilla, Europhys.\ Lett.\ {\bf 45}, 700 (1999).

\end{thebibliography}
\end{document}